\documentclass[aps,preprint,epsfig,rotate,superscriptaddress,showpacs]{revtex4}
\usepackage{epsfig}
\usepackage{graphicx}
\usepackage{amsmath}

\newcommand{\beq}{\begin{eqnarray}}
\newcommand{\eeq}{\end{eqnarray}}
\begin{document}

\title{Thermal conductance in thin wires with surface disorder}
\author{Gursoy B. Akguc} \affiliation{Department
of Physics and Centre for Computational Science and Engineering, \\
National University of Singapore, 117542, Republic of Singapore}
\author{Jiangbin Gong} \email{phygj@nus.edu.sg}
\affiliation{Department
of Physics and Centre for Computational Science and Engineering, \\
National University of Singapore, 117542, Republic of Singapore}
\affiliation{NUS Graduate School for Integrative Sciences and
Engineering, Singapore
 117597, Republic of Singapore}

\begin{abstract}
Elastic wave characteristics of the heat conduction in
low-temperature thin wires can be studied via a wave scattering
formalism. A reaction matrix formulation of heat conductance modeled
by elastic wave scattering is advocated. This formulation allows us
to treat thin wires with arbitrary surface disorder. It is found
that the correlation in the surface disorder may significantly
affect the temperature dependence of the heat conductance.
\end{abstract}

\date{\today}
\maketitle
\section{Introduction}

In a thin wire where the electron or phonon wavelength is in the
same order of its width,  a continuous description of transport is
no longer valid. Instead, a quantized unit of electric charge $e$ or
a characteristic scale of heat energy $k_B T$ ($k_B$ is Boltzmann
constant and $T$ is the temperature) is needed to understand the
transport properties.  When such a wire is under an external field,
e.g., a voltage difference in the electron case or a temperature
difference in the phonon case, system-independent features such as
the universal electric conductance quanta $G_e=2e^2/\hbar$ ($\hbar$
is the Planck constant) and the universal heat conductance
$G=4\pi^2k_b^2T/3\hbar$ quanta will emerge \cite{Luis}. With the
universal heat conductance quanta successfully demonstrated recently
\cite{Schwab}, there have been great interests in the wave nature of
heat transport at the nanoscale.  For example, recent experiments
showed that the thermoelectric properties of silicon nanowires are
much better than that in bulk silicon \cite{Nature}.

Besides a nanoscale geometry, one needs low temperatures (e.g., $T
\sim 1$ K) to observe a phase-coherent transport and hence the wave
nature of phonon transport \cite{Schwab,SYLi,Li}. At low
temperatures only acoustic phonon modes are populated, with their
characteristic wavelength much longer than typical atom-atom
distances. As such, an elastic wave approach becomes appealing for a
quasi-classical description of heat transport.  Indeed, elastic wave
propagation in thin wires has been analyzed in several studies with
fruitful results \cite{Wen1, Wen2, Wang, Hai, Lim}. However, in
these earlier publications only very simple geometries of a thin
wire were considered, leaving the case of a thin wire with surface
disorder unexplored. This motivates us to develop a framework that
can handle elastic wave scattering for arbitrary wire geometries.
Specifically, we apply a reaction matrix formulation, previously
thoroughly developed for electron transport in nanoscale waveguides
\cite{akg1}, to the case of heat transport. We believe that this is
the first time that a reaction matrix formulation is applied to the
context of elastic wave scattering as well as heat conductance. In
so doing we use the so-called thin-plate approximation, which means
that one of the dimensions is decoupled from the other two. This
thin-plate approximation is widely used in analyzing transport
behaviors \cite{Cross1}.

Our formalism here for treating elastic wave transport can handle
arbitrary surface geometry or arbitrary surface disorder. This
feature can be very useful because, depending on the process of
growth, there are often different types of surface disorder
introduced to nanowires \cite{Chen}. Preliminary studies on the
effect of surface disorder on nanowire heat conduction show the
existence of universal features in the presence of disorder
\cite{Glavin,Cross2,Murphy}. Nevertheless, it is still an important
problem to understand in detail how surface disorder affects the
actual temperature dependence of heat conduction.
Furthermore, as learned from our previous work on electron transport
in rough waveguides, conductance properties of a nanoscale rough
waveguide can depend strongly on the involved energy scale and the
long-range correlation of the surface disorder.  We hence expect,
similar to early studies of heat transport in one-dimensional
systems with partially random defects or partially random coupling
constants \cite{Ke,Cao, Abh},
that long-range correlation in the surface
disorder of thin wires may have important implications for heat
transport properties.  

In addition to presenting details of a reaction matrix formulation
of elastic wave scattering in two-dimensional geometries, we also
discuss several specific results. For example, it is found that the
throughput can be indeed modified by assuming different kinds of
correlations in the surface disorder of a wire. In the one-mode
regime the curve of transmission versus phonon frequency may develop
a clear dip, thus considerably affecting the temperature dependence
of the heat conduction.

This work is organized as follows. In Sec. II  we introduce our
elastic wave scattering model for heat transfer in long thin wires.
In Sec. III, we first introduce how specific surface roughness can
be generated, and then show the details of how the throughput of
out-of-plane elastic waves can be calculated using a reaction matrix
formulation. Representative numerical results are presented and
discussed. Section IV concludes this work.

\section{Heat transport modeled as a problem of two-dimensional elastic wave scattering}

\begin{figure}
\includegraphics[width=14cm]{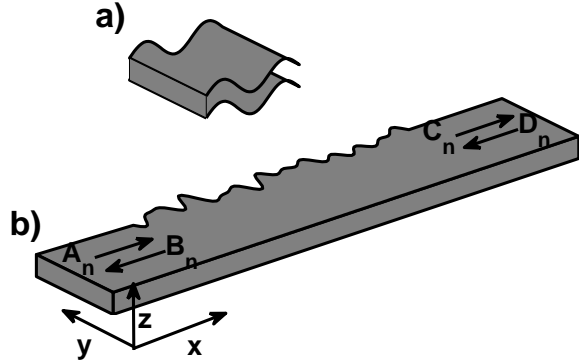}
\caption{(a) Schematic plot of a out-of-plane mode of an elastic
wave propagating in a thin plate. (b) Schematic plot of a
two-dimensional rough waveguide that models a thin wire with a rough
surface at $y>0$. $A_n$, $B_n$, $C_n$ and $D_n$ denote scattering
amplitudes. See the text for details.}
\end{figure}

Heat transport in a solid is a result of unbalanced phonon
population. There are mainly two different perspectives in
understanding the heat current in nano-materials \cite{Mass}. One is
based on the Kubo formalism, where phonon current is understood as a
response to an external temperature field, with the proportionality
coefficient expressed as a heat conductivity tensor. Similar to the
electron transport case, the conductivity tensor is related to the
current-current correlation function. Because the current-current
correlation function should be calculated for different phonons and
because the phonon number can change (um-klapp process), the Kubo
formalism leads to nonlinear equations that are difficult to solve
in practice. The other perspective is due to Landauer, where the
current can be determined at the surfaces of the sample by applying
the correct boundary conditions. In particular, the known solution
for outside a sample can be matched at the boundary using a
scattering formalism. In this latter framework, from the right bath
with temperature $T_R$ to the left bath with temperature $T_L$, the
heat energy flux $J$ is given by
\begin{equation}
J=\frac{1}{2\pi}\sum_\alpha\int_0^\infty dw\ \hbar
w(\eta_R-\eta_L)G_\alpha(w), \label{eq1}
\end{equation}
where $w$ is the phonon frequency, $G_\alpha$ is the throughput of
the mode $\alpha$. Here $\eta_{R,L}$ is the phonon distribution
given by $\eta_i(w)=1/[\exp(\hbar w/k_B T_i)-1]$, where $T_i$ are
the temperatures, with indices $i=R,L$. Heat conductance is defined
by energy flux  per temperature difference, i.e., \begin{eqnarray}
 \kappa=J/\Delta T\approx \partial J/\partial T,
\end{eqnarray}
where for small temperature difference ($T_{R}-T_{L})$, $\kappa$ is
approximated by the temperature derivative of the energy flux. After
substituting $\eta_i(w)$ into Eq. (\ref{eq1}) we have,
\begin{equation}
\kappa=\frac{\hbar^2}{k_BT^2}\sum_\alpha\frac{1}{2\pi}\int_0^\infty
dw G_\alpha(w) \frac{w^2\exp(\hbar w/k_BT)}{[\exp(\hbar
w/k_BT)-1]^2}, \label{eq:he}
\end{equation}
where we have defined the average temperature $T\equiv(T_L+T_R)/2$.
In a perfect wire throughput is unity for each open channel, then
$\kappa$ in such a perfect case is given by
\begin{equation}
\kappa=\frac{k_B^2 T \pi^2}{3\hbar} N,
\end{equation}
where $N$ is the total number of open modes.  As seen from Eq.
(\ref{eq:he}), the throughput $G_\alpha(w)$ is the main quantity
that determines the temperature dependence of $\kappa$ in Landauer's
formalism.  In the language of a scattering problem, $G_\alpha(w)$
is determined by the absolute value squared of certain scattering
amplitudes.



To calculate the throughput of elastic waves in a thin wire, we use
a thin plate approximation to treat full elastic wave equations. A
schematic plot of a two-dimensional rough waveguide modeling a
nanowire with surface disorder is shown in Fig. 1(b).   Because at
low temperatures contribution from any optical modes is
exponentially small \cite{Luis},  we use an elastic theory to
describe elastic wave scattering in a relatively long wire.  As we
show in Fig.~1(b) as an example, we intend to solve a scattering
problem in a wire with the width-length ratio set to be 1:100. We
scale all the lengths by the wire width, which is taken to be unity.

The elastic wave equation for our model system is given by
\begin{equation}
\frac{\partial^2 {\bf u}}{\partial t^2}=c_t^2\nabla^2{\bf u} +
(c_l^2-c_t^2)\nabla(\nabla\cdot {\bf u}), \label{eq:elas}
\end{equation}
where ${\bf u}$ is the displacement vector , $c_l$ is the
longitudinal phonon velocity and $c_t$ is the transverse phonon
velocity \cite{Graff}.  For convenience we also scale the velocities
by $c_t$.  That is, in our calculation $c_t=1$.  In the thin-plate
approximation,  the gradient in the $z$-direction is assumed to be
zero, thus simplifying Eq.~(\ref{eq:elas}). Such an approximation
decouples the in-plane acoustic modes from the out-of-plane acoustic
modes. Assuming that the in-plane and out-of-plane modes behave in a
similar fashion, we confine ourselves to the out-of-plane modes only
[see Fig. 1(a)].  Technically speaking, implementing the reaction
matrix formulation for out-of-plane modes is also simpler than for
in-plane modes.  The wave equation for out-of-plane modes becomes
\begin{equation}
\frac{1}{c_t^2}\frac{\partial^2 u_z}{\partial t^2}=\frac{\partial^2
u_z}{\partial x^2}+ \frac{\partial^2 u_z}{\partial y^2},
\label{eq:oop}
\end{equation}
where $u_z(x)$ is the displacement in the $z$ direction. The
boundary condition for this wave equation will be specified below.

The next step is to calculate the scattering of the elastic waves in
a two-dimensional geometry as shown in Fig. 1(b).  We will employ a
reaction matrix formulation for this scattering problem. The basic
idea of a reaction matrix approach is as follows. First, the system
is divided into the lead region and the scattering region. Second,
the known solution in the lead region is matched at the boundary
with the found solution in the scattering region. This matching is
implemented by using a basis set satisfying the appropriate boundary
condition.  In the following section we will show how the reaction
matrix can be constructed based on appropriate basis states.

\section{Scattering throughput in a thin wire with surface disorder}

As learned from electron scattering in a rough waveguide
\cite{akg2}, it can be expected that disorder-induced localization
will also play an important role in elastic wave scattering in a
thin wire.
Furthermore,  if the localization length is smaller than the wire
length, then the scattering throughput $G_{\alpha}(w)$ is
exponentially small; if the localization length is much larger than
the wire length, then $G_{\alpha}(w)$ should be close to unity.
Because in the low-temperature regime only a few scattering channels
contribute to heat transport, it is also important to realize that
the localization length might depend strongly on the phonon
frequency \cite{Murphy}. In particular, based on the Born
approximation, the electron localization length is inversely
proportional to the structure factor (defined below) of the
correlation function of the surface disorder evaluated at twice of
the scattering electron wavevector \cite{Izr,akg2}. By direct
analogy between the Schr\"{o}dinger equation for electron matter
waves and the elastic wave of matter, we can expect that the
correlation function of thin-wire surface disorder can also affect
strongly the localization length of a scattering elastic wave.
Therefore, we are interested in (i) if it is possible to manipulate
the frequency dependence of the scattering throughput of elastic
waves, and (ii) to what extent different surface disorders might
affect the heat conductance.


For completeness we first discuss how to computationally produce a
random surface characterized by $y=1+\eta(x)$ [see Fig. 1(b)]. This
can be done by dividing a long wire into pieces, shifting each piece
up or down randomly, and then connecting each piece smoothly with a
cubic spline function \cite{akg2,prb1}. Randomness of the surface
function $\eta$ thus generated can be described by an
auto-correlation function
$\langle\eta(x)\eta(x')\rangle=\sigma^2C_n(x-x')$, where $\sigma$ is
the variance of $\eta(x)$. The Fourier transform of this auto
correlation function is defined as the structure factor, denoted
$\chi(k)$. For disorder close to a white-noise type or for disorder
without  long-range correction (computationally, we can only divide
the wire region into a finite number of pieces, e.g., 100 pieces,
hence the surface disorder we generate is at most close to a
white-noise type), the structure factor trivially takes a constant
value over a broad range of $k$.
 To
generate a more realistic rough thin wire  with ``colored" or
long-range surface disorder,  we define another function $\rho(x)$,
\begin{equation}
\rho(x)=\frac{\sin(|a^r|x)-\sin(|a^l| x)}{|a^r|x}
\end{equation}
where $a^l$ and $a^r$ are two introduced parameters. Consider then a
new function $\tilde{\eta}(x)$ resulting from a convolution of
$\eta(x)$ with $\rho(x)$, i.e., $\tilde{\eta}(x)=\int
dx'\rho(x')\eta(x-x')$. It can be easily shown \cite{akg2} that due
to this convolution, $\chi(k)$ now becomes a ``square bump"
function, with the bump edges located at $a^l$ and $a^r$. That is,
$\chi(k)$ is zero for $k<a^l$ or $k>a^r$ and is a constant in
between. This convolution method is used below to produce several
``bump" functions as an ``engineered"  surface-disorder structure
factor, with different bump widths.  Certainly, one may also combine
a number of different $\rho(x)$ functions to generate a rather
arbitrary structure factor $\chi(k)$.



\subsection{Constructing the reaction matrix}
The reaction matrix method is a well-known time-independent
scattering formalism. As such we need to consider the stationary
solutions to the wave equation in Eq. (\ref{eq:oop}).  Let us first
substitute the ansatz $\exp(-iwt)u_z$ into Eq. (\ref{eq:oop})
\cite{Graff}, yielding
\begin{equation}
\frac{\partial^2 u_z}{\partial x^2}+\frac{\partial^2 u_z}{\partial
y^2}-\frac{w^2}{c_t^2}u_z=0, \label{tieq2}
\end{equation}
where $w$ is the phonon frequency. 
The realistic boundary condition for $u_z$ is given by \cite{Graff}
\begin{equation} \frac{\partial u_z}{\partial y}=0.
\end{equation}
A stationary solution to Eq. (\ref{tieq2}) for a straight channel
satisfying the above ``zero-stress" boundary condition at $y=0$ and
$y=1$ can then be found, namely,
\begin{eqnarray}
\Psi_{n \emph{l}}&=&  \left( A_n \frac{e^{ik_nx}}{\sqrt{k_n}}
-B_n\frac{e^{-ik_nx}}{\sqrt{k_n}}\right)
\cos(\frac{n\pi y}{{\it d}}); \label{lead1}  \\
\Psi_{n \emph{r}}&=& \left(C_n \frac{e^{ik_nx}}{\sqrt{k_n}}
-D_n\frac{e^{-ik_n x}}{\sqrt{k_n}}\right) \cos(\frac{n\pi y}{{\it
d}}), \label {lead2}
\end{eqnarray}
where $A_n$, $B_n$, $C_n$, and $D_n$ are the scattering amplitudes.
The wavevector is given by $k_n=\sqrt{w^2-(n\pi/{\it d})^2}$, where
$n=0,1,2, \cdots$. Note that when $k_n$ takes imaginary values, the
solutions are called ``evanescent modes". As mentioned above, in our
unit system $d=1$, the wire length is $L=100$, and the phonon
frequency $w$ is in terms of the value of the first mode, i.e.,
$w_1=c_t \pi/ d=\pi$. We also scale temperature by $T_D$, where
$T_D\equiv \hbar w_1/k_B$.  For a thin wire of a width $d=10^{-6}$cm
and for an acoustic transverse phonon speed (Silicon) $c_t=5.84
10^5$ cm/sec,
we have $T_D=14$K. 

In a reaction matrix formulation of wave scattering, solutions in
Eqs. (\ref{lead1}) and (\ref{lead2}) can be taken as ``free"
solutions in the left and right leads that are connected by the
wire. To find the relation between the wave amplitudes $A_{n}$,
$B_{n}$, $C_{n}$, and $D_{n}$, these two solutions will be matched
with that in the wire. Hence, we need to solve the elastic wave
equation in Eq. (\ref{tieq2}), confined by two boundaries defined by
$y=1+\tilde{\eta}(x)$ (or $y=1+{\eta}(x)$ for an almost white-noise
type surface disorder) and $y=0$ [see Fig.~1(b)]. To that end a set
of eigenfunctions of Eq. (\ref{tieq2}) satisfying the zero-stress
boundary condition are needed. In addition, as one main feature of
the reaction matrix formulation \cite{akg1,akg2}, the derivative of
the eigenfunctions with respect to the scattering coordinate $x$
should be zero at the lead-wire interface, i.e., at $x=0$ and $x=L$.

To find the eigenfunctions in the scattering region, we employ a
coordinate-transformation technique.  That is, we consider two new
coordinates $x'=x$ and $y'=y/[1+\tilde{\eta}(x)]$, where
$\tilde{\eta}(x)$ is the surface function of the wire. In the
$x'-y'$ coordinate system, the elastic wave equation becomes
complicated. However, it is now considered over a straight channel
(see Refs. \cite{akg1,akg2} for details of this
coordinate-transformation technique).
In the new coordinate system an eigenstate $\phi_n$ satisfying the
necessary boundary condition can be written as
\begin{equation}
\phi^i(x',y')=\sum_{m,m'}\beta_{m,m'}^i\frac{2}{\sqrt{L(1+\tilde{\eta}(x')}}\cos(\frac{m\pi
x'}{L})\cos(m'\pi y'),
\end{equation}
where $m,m'$ are integers,  the summation over $m$ and $m'$ must be
truncated for practical purposes, and $\beta^{i}_{mm'}$ are the
expansion coefficients to be found numerically.

Once the eigenstates $\phi^i(x',y')$ with eigenfrequencies
$\tilde{w}_{i}$ are obtained, we impose the continuity condition at
$x=0$ and $x=L$, yielding a reaction matrix,
\begin{equation}
R_{n m}=\sum_{i=1}^{M} \frac{\phi_n^i(0)\phi_m^i({\it
l})}{w-\tilde{w}_i},
\end{equation}
where $M$ is the size of the involved basis sets (exact for
$M=\infty$), and $\phi_n^i=\int_0^1{\phi^i\cos(n\pi y)}$ is the
overlap of $i$th eigenstate with the $n$th ``channel" solution
$\cos(n\pi y)$.  The index $n,m$ of the reaction matrix $R_{nm}$
denote the scattering channels associated with real wavevectors
$k_{n}$ and $k_{m}$. The reaction matrix $R_{nm}$ can then yield the
scattering matrix,
\begin{equation}
S=\frac{I-iKR}{I+iKR},
\end{equation}
where $I$ is a unit matrix of the same size with $R$, and $K$ is a
diagonal matrix with diagonal elements related to the wavevectors
$k_n$ \cite{akg2,prb1}.  The scattering matrix then relates the
incoming wave with the outgoing wave via
\begin{equation}
\left( \begin{array}{c} B_n \\ C_n \end{array}\right)
=S\left(\begin{array}{c} A_n \\ D_n \end{array} \right).
\end{equation}
Finally, the total scattering throughput is defined as $G\equiv
\sum_{\alpha}G_{\alpha}=\sum_{n,m} |S_{n,m}|^2$, where the summation
is over all open modes.  $G$ essentially describes the total
transmission of a scattering wave up to a certain given phonon
frequency, without taking into account the actual thermal
distribution of these phonon modes. 


\begin{figure}
\includegraphics[width=10cm]{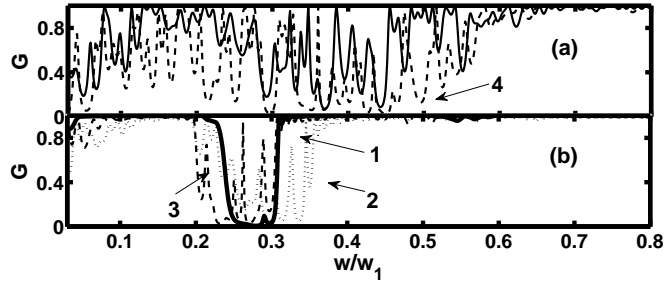}
\caption{Acoustic phonon transmission in the one-mode regime, as
reflected in the dimensionless throughput $G$. (a) Results for two
realizations of a waveguide with almost white-noise surface
disorder. (b) Results for three different cases of colored surface
disorder, with the associated structure factor set as a ``square
bump" function with three different bump widths. See the text for
details.}
\end{figure}
\subsection{Throughput and heat-conductance results}
At low temperatures, very few acoustic modes can be populated. Hence
it is of interest to examine the scattering throughput in the
one-mode regime. This is shown in Fig. 2. Specifically, Fig.~2(a)
depicts the results associated with two realizations of a wire with
almost white-noise surface disorder. It is seen that when the phonon
frequency is small, the throughput $G$ is fluctuating drastically,
with its average value far from unity. This indicates that in these
two cases the localization length of the elastic wave is comparable
to the system size.  As the phonon frequency increases, $G$ is seen
to be close to unity, suggesting that the localization length is now
well beyond the wire length. Figure 2(b) shows what happens if we
impose a correlation on the surface disorder, but with the noise
variance fixed at $\sigma=0.08$.  In particular, the dotted curve
(curve 2), the dashed curve (curve 3), and the solid curve (curve 1)
are for ``engineered" surface functions generated from a convolution
of a random surface function $\eta(x)$ with $[\sin(0.7\pi
x)-\sin(0.4\pi x)]/(0.7\pi x)$, $[\sin(0.7\pi x)-\sin(0.5\pi
x)]/(0.7\pi x)$, and $[\sin(0.6\pi x)-\sin(0.5\pi x)]/(0.6\pi x)$,
respectively.   It is seen that $G$ develops a clear window for
certain phonon frequencies.   Despite statistical fluctuations,
these throughput windows also show differences in their widths,
namely, curve 2 shows the widest window and curve 1 shows the
narrowest window. This is consistent with the differences in the
bump widths of their associated surface structure factor.
We also examined other realizations of random surfaces with the same surface structure factor,
confirming that the location of the throughput windows shown in Fig. 2
does not change much.
We have
also checked that the location of the throughput window
quantitatively matches the profile of $\chi(2k)$, the surface
structure factor evaluated at twice of the wavevector.

\begin{figure}
\includegraphics[width=10cm]{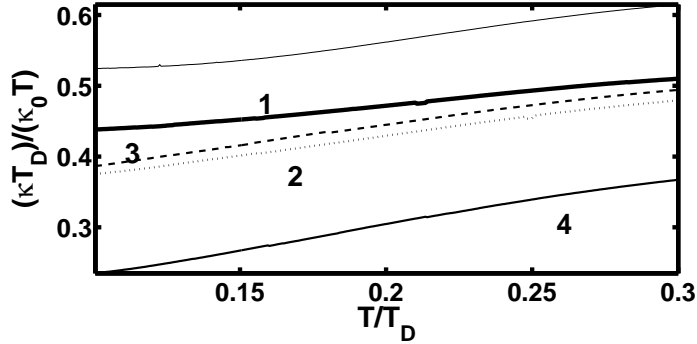}
\caption{The ratio between heat conductance $\kappa$ and
temperature, as a function of temperature,  for wire geometries
considered in Fig.~2.  Thin solid line is for a wire without surface
disorder. Line (4) is for a wire with almost white-noise surface
disorder. Other lines are for wires with their surface function
generated from a convolution of white-noise with some smooth
function explicitly given in the text. Here and in all other figures
below, the plotted quantity is always dimensionless and
$\kappa_0\equiv k_B w_1$.}
\end{figure}

In Fig.~3 we examine how different properties of the wire disorder
might affect the heat conductance $\kappa$ calculated from Eq. (3).
To make a better connection with the results in Fig. 2 we examine a
temperature regime where thermal excitation only populates at most
two channels.  First of all, for the case without disorder (upper
thin solid curve), the plotted quantity $\kappa/T$ is seen to be
roughly a constant for low temperatures. This is somewhat expected
because a perfect thin wire is known to show a linear temperature
dependence in $\kappa$ at low temperatures. Certainly, as the
temperature increases, deviations from the linear behavior emerge
due to an increasing contribution from the second populated channel.
By contrast, in the presence of surface disorder, the linear
temperature dependence of $\kappa$ might not hold from the very
beginning of the temperature range shown in Fig. 3 (curves 2,3,4).
Interestingly, for different surface disorder, $\kappa/T$ shows
quite different behavior. The almost white-noise case (curve 4)
deviates most from the case without disorder, with the maximal
deviation around a factor of two. These observations suggest that
one might achieve some subtle control over the temperature
dependence of $\kappa$ in the low-temperature regime by manipulating
surface correlations,  or say something about the surface
correlation by measuring $\kappa$ at different temperatures.

\begin{figure}
\includegraphics[width=10cm]{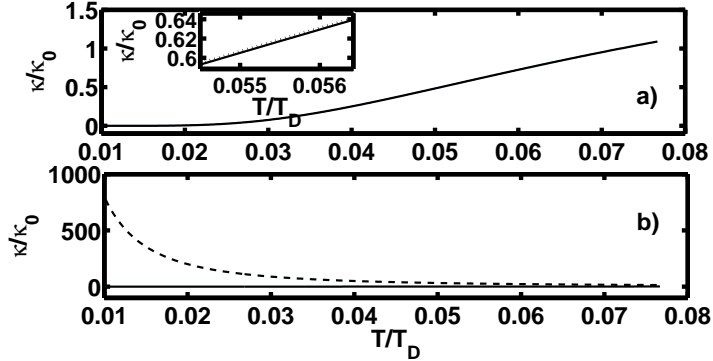}
\caption{(a) Heat conductance from direct numerical integration
(dotted) of Eq. (3) versus the analytical result (solid line) of
Eq.~(\ref{eq:hc}), for $a=0.2$. As indicated by the inset, the
dotted line is on top of the solid line. (b) Same as in (a) but the
dashed line is calculated from an expression in Ref. \cite{Luis}.
Note the different scales between panel (a) and panel (b). }
\end{figure}

To shed more light on how the throughput windows observed in Fig.
2(b) affect the temperature dependence of $\kappa$, let us now
consider a model throughput curve that allows for analytical
calculations.  Consider first two step functions $f_a(w)$ and
$f_{b}(w)$, namely, $f_a(w)=0$ for $w<a$ and $f_a(w)=1$ for $w>a$;
and $f_b(w)=0$ for $w<b$ and $f_b(w)=1$ for $w>b$. We denote the
heat conductance as $\aleph(a)$ for $G(w)=f_a(w)$ and $\aleph(b)$
for $G(w)=f_b(w)$, respectively. Assuming that there is only one
open mode,  we obtain from Eq. (3) that
\begin{equation}
\aleph(a)=k_Bw_1\left(\frac{a^2e^{\frac{a}{T}}}{T(e^{\frac{a}{T}}-1)}
-2 \log \left(e^{a/T}-1\right) a+\frac{2 \pi ^2 T}{3}-2 T
\Re\left[\text{Li}_2\left(e^{a/T}\right)\right]\right) \label{eq:hc}
\end{equation}
and an analogous expression for $\aleph(b)$,  where $Li_{2}$ is the
so-called dilog function. 
 Interestingly, at this point we found
that the widely cited Ref. \cite{Luis} does not contain the second
term in Eq. (\ref{eq:hc}) and misses a factor of two in the third
term.  To ensure that our expression is indeed correct we show in
Fig. 4(a) that direct numerical integration results based on Eq.
(\ref{eq:he}) are identical with the analytical result of Eq.
(\ref{eq:hc}) for $a=0.2$. For the sake of comparison we also show
in Fig.~4(b) the associated analytical result of Ref. \cite{Luis}
(dashed line), as compared with the same solid line in Fig. 4(a).
Note also that because the function $Li_{2}(z) $ for certain values of $z$ can be
explicitly evaluated, it is now possible to predict the exact value of $\aleph(a)$
for certain $e^{a/T}$.

Let us now examine the heat conductance behavior if $G(\omega)$ is a
square-well function $G_{sw}(w)$, i.e., it is zero for $a<w<b$ and
is unity elsewhere. Because $G_{sw}(w)=f_{b}(w)+1-f_{a}(w)$, one
obtains from Eq. (\ref{eq:hc})
\begin{equation}
\kappa=\frac{k_Bw_1 \pi^2T}{3}+\aleph(b)-\aleph(a), \label{generalB}
\end{equation}
where $T$ is already scaled by $T_D$. Approximately, for $b>a>>T$,
we have
\begin{equation}
\kappa=\frac{k_Bw_1 \pi^2T}{3}+k_Bw_1\left(\frac{T^2(b-a)}{(T-a)(T-b)}\right);
\label{lowT}
\end{equation}
and for $a<b<<T$  we have
\begin{equation}
\kappa=\frac{k_Bw_1
\pi^2T}{3}+k_Bw_1\left(a-b+2a\log(a)-2b\log(b)+\frac{b^3-a^3}{36T^2}\right).
\label{bigT}
\end{equation}
 These two limiting results show that in either case [Eqs.
(\ref{lowT}) or Eq.
(\ref{bigT})], a window in the throughput curve can cause $\kappa/T$
to decrease as compared with what is expected from a linear
temperature dependence of $\kappa$.

\begin{figure}
\includegraphics[width=10cm]{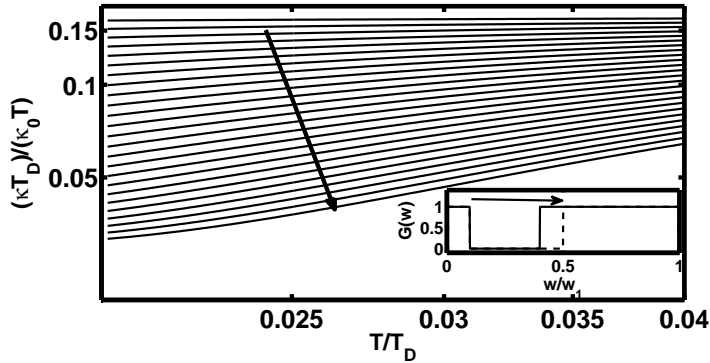}
\caption{Temperature dependence of $\kappa/T$, for throughput curves
modeled by square-well functions. The arrow indicates an increasing
width of the square-well function, in steps of  0.08, ranging from
0.0 to 0.4. Note that quantities here are plotted in the logarithmic
scale.}
\end{figure}

From Eq. (\ref{generalB}) it is also possible to examine how the
width $(b-a)$ of the square well function $G_{sw}(w)$ as a
throughput curve affects $\kappa/T$. To that end we show in Fig. 5 a
logarithmic plot of $\kappa/T$ versus $T/T_{D}$. The arrow therein
indicates an increasing $(b-a)$ in steps of 0.08, which ranges from
0.0 to 0.4. Interestingly, as the throughput window increases its
size, $\kappa/T$ decreases.  The $\kappa/T$ function is also seen to
deviate from a constant function more and more, and a better
description of the temperature dependence of $\kappa$ seems to be a
power function. We note that this observation is consistent with
experimental observations in superconductive materials \cite{SYLi}.
Note also that the analytical result here also explains the general
trend seen in our numerical results in Fig. 3. That is, the lowering
of $\kappa/T$ curves in Fig. 3 is associated with the widening of
the throughput windows shown in Fig. 2(b).

So far we have studied the low-temperature regime where one or two
modes are significantly populated.  What happens if more phonon
modes are populated?  To answer this question we show in Fig. 6 the
throughput $G(w)$ for up to five modes, for a wire without disorder
[case (a)], with almost white-noise surface disorder [case (b)], and
with colored surface disorder [cases (c) and (d)].  A number of
observations are in order. First, $G(w)$ can be nonzero for phonon
frequencies much smaller than $w_1$, a feature different from that
in electron transport. In the latter case one can only excite the
first mode after going beyond the associated cut-off energy. This
difference arises from the fact that the Schr\"{o}dinger equation
for electrons and the elastic wave equation for phonons have
different boundary conditions.  Second, in the presence of surface
disorder,  the threshold frequency for opening a new channel and
hence a drastic increase in $G(w)$ gets higher as compared with the
noiseless case. This is analogous to the electron case \cite{akg2},
which can be roughly explained via a decrease in the effective wire
width. Third, for cases (c) and (d), the dip in $G(w)$ is clear only
for the one-mode and two-mode regimes. For higher modes, the
throughput dip is no longer clear and is buried by fluctuations in
$G(w)$.

\begin{figure}
\includegraphics[width=10cm]{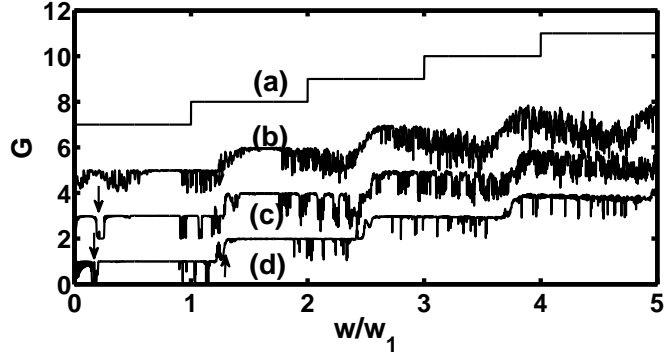}
\caption{Throughput $G(w)$ versus phonon frequency. Case (a) is for
a perfect wire, case (b) is a wire with white-noise surface
disorder, and cases (c) and (d) are for two realizations of a wire
with colored surface disorder. For a better view cases (c), (b), (a)
are shifted upwards by two, four, and six units, respectively. The
arrows indicate the dip in the throughput curves. For example, the
right arrow for curve (d) indicate a throughput dip in the two-mode
regime.}
\end{figure}

Can those throughput curves in Fig. 6 manifest themselves
differently in the heat conductance behavior, when the temperature
is high enough to populate many phonon modes? The answer is positive
based on the results shown in Fig. 7.  In particular, Fig. 7 depicts
the temperature-dependence of $\kappa$ for those cases shown in Fig.
6. It is seen that these cases still show quite different behavior
even when five phonon modes are significantly populated. Cases (c)
and (d) associated with colored surface disorder lie between the
curve for a perfect wire and that for a wire with almost white-noise
surface disorder.  Results in Fig. 7 indicate that for temperatures
of the order of $\sim$ 10 K, different surface disorder properties
of a thin wire of a width $\sim$ 10 nm may still cause a
considerable difference in heat conductance.  The results are also plotted on a log-log scale in the inset of Fig. 7.
However, we do not see any evident
power-law dependence of $\kappa$ versus $T$. This is somewhat expected because the T-dependence of $\kappa$
is quite complicated [see, for example, Eq. (17)]. 
For the sake of
comparison, a perfect wire with twice the width is also shown in
Fig. 7 as case (e). In that case the temperature
dependence is much stronger. This is anticipated because of the $T^3$ scaling of
$\kappa$ in bulk materials.  

\begin{figure}
\includegraphics[width=10cm]{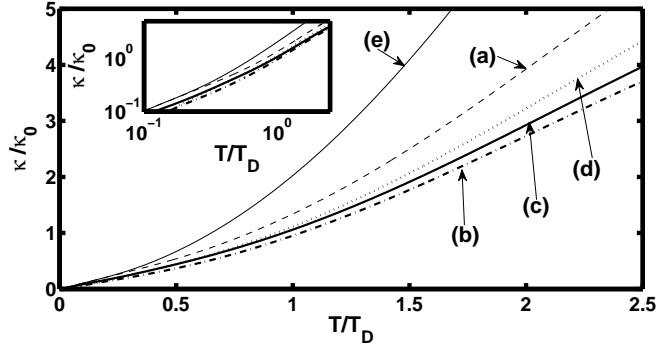}
\caption{Temperature-dependence of heat conductance for those cases
considered in Fig.~6.  Case (e) is for a perfect (straight) wire
with twice the width, in which the temperature dependence of
$\kappa$ is approaching the $T^3$ law for bulk materials. Inset shows the same data in a
log-log plot. }
\end{figure}

\section{Concluding remarks}
In this work we adopted an elastic wave picture of heat transport at
sufficiently low temperatures. We presented, for the first time, a
reaction matrix treatment of the scattering of elastic waves in thin
wires with surface disorder, with realistic zero-stress boundary
conditions implemented.   We have shown that correlations in the
surface disorder can lead to structures in the frequency-dependence
of throughput and can hence affect considerably the temperature
dependence of heat conductance. Though not done in this study,   the
heat conductance for a thin wire with an experiment-related surface
correlation function can be examined in a straightforward manner.
Because our treatment is a natural extension of early electron
transport studies, insights gained from studies of electron
waveguide transport can now be much relevant to understanding heat
transport in thin wires.

Because the boundary condition for in-plane modes of elastic waves
is much more difficult to handle for a thin wire of arbitrary shape,
here we restrict ourselves to out-of-plane modes only. However,
preliminary results show that in-plane modes have similar
qualitative behavior as out-of-plane modes.

\acknowledgments We thank Prof. Li Baowen for helpful discussions.
This work was supported by the start-up fund (WBS grant No.
R-144-050-193-101/133) and the NUS ``YIA" fund (WBS grant No.
R-144-000-195-101), both from the National University of Singapore.
We also thank the Supercomputing and Visualization Unit (SVU), the
National University of Singapore Computer Center, for use of their
computer facilities.

\


\begin{thebibliography}{10}

\bibitem{Luis} 
L. G. C. Rego and G. Kirczenow, Phys. Rev Lett. {\bf 81},  232
(1998).

\bibitem{Schwab} 
K. Schwab, E. A. Henriksen, J. M. Worlock, and M. L. Roukes, Nature
(London) {\bf 404}, 974 (2000).


\bibitem{Nature} 
A. I. Hochbaum, R. K. Chen, R. D. Delgado, W. J. Liang, E. C.
Garnett, M. Najarian, A. Majumdar, and P. D. Yang, Nature, {\bf 451}
No. 10, 163, (2008).

\bibitem{SYLi} 
S. Y. Li, J.-B. Bonnemaison, A. Payeur, P. Fournier, C. H. Wang, X.
H. Chen, and L. Taillefer, Phy. Rev B. {\bf 77}, 134501 (2008).

\bibitem{Li} 
D. Y. Li, Y. Y. Wu, P. Kim, P. D. Yang, and A. Majumdar,
Appl. Phys. Lett. {\bf 83}, 2934 (2003).


\bibitem{Wen1} 
W. X. Li, K. Q. Chen, W. H. Duan, J. Wu, and B. L. Gu, J. Phys.:
Condens. Matter {\bf 16}, 5049 (2004).


\bibitem{Wen2} 
W. X. Li, T. Y. Liu, C. L. Liu, Chin Phys. Lett. {\bf 23}, 2522
(2006).


\bibitem{Wang} 
X. F. Wang, M. S. Kushwaha, and P. Vasilopoulos,
Phy. Rev B. {\bf 65}, 035107 (2001).


\bibitem{Hai} 
H. Y. Zhang, H. J. Li, W. Q. Huang, and S. X. Xie, J. Phys. D: Appl.
Phys. {\bf 40}, 6105 (2007).

\bibitem{Lim} 
L. M. Tang, L. L. Wang, W. Q. Huang, B. S. Zou, and K. Q. Chen, J.
Phys. D: Appl. Phys. {\bf 40}, 1497 (2007).


\bibitem{akg1} 
G. B. Akguc and T. H. Seligman, Phys. Rev. B. {\bf 74}, 245317
(2006).


\bibitem{Cross1} 
M. C. Cross, R. Lifshitz, Phy. Rev B. {\bf 64}, 085324 (2001).

\bibitem{Chen} 
R.K. Chen, A. I. Hochbaum, P. Murphy, J. Moore, P.D. Yang, and A.
Majumdar, Phy. Rev Lett. {\bf 101}, 105501 (2008).

%

\bibitem{Glavin} 
B. A. Glavin, Phy. Rev Lett. {\bf 86}, 4318 (2001).


\bibitem{Cross2} 
D. H. Santamore and M. C. Cross,
Phy. Rev B. {\bf 66}, 144302 (2002).


\bibitem{Murphy} 
P. G. Murphy and J. E. Moore, Phy. Rev B. {\bf 76}, 155313 (2007).

\bibitem{Ke} 
K. Q. Chen and X. H. Wang, and B. Y. Gu, Phy. Rev B. {\bf 61}, 12075
(2000).


\bibitem{Cao} 
L. S. Cao, R. W Peng, R. L. Zhang, X. F. Zhang, M. Wang, X. Q.
Huang, A. Hu, and S. S. Jiang, Phy. Rev B. {\bf 72}, 214301 (2005).


\bibitem{Abh} 
A. Dhar, Phy. Rev. Lett. {\bf 86}, 5882 (2001).

\bibitem{Mass} 
{\it Electric Transport in Nanoscale Systems}, Massimiliano Di
Ventra, Cambridge, 2008.

\bibitem {Graff}
{\it Wave Motion in Elastic solids}, K. F. Graff, Clarendon Press,
Oxford, 1975;\ {\it Introduction to Elastic Wave Propagation}, A
Bedford and D. S. Drumheller, John Wiley-Sons, 1994;\  {\it Theory
of Elasticity, 3rd Edition}, L.D. Landau and E.M. Lifshitz, Elsevier
1986.

\bibitem{akg2} 
G. B. Akguc and J.B. Gong, Phys. Rev. B. {\bf 78}, 115317 (2008).

\bibitem{Izr} 
F. M. Izrailev and A. A. Krokhin, Phy. Rev. Lett. {\bf 82}, 4062
(1999).

\bibitem{prb1}
G. B. Akguc and J.B. Gong, Phys. Rev. B. {\bf  77}, 205302 (2008).




%
%
%
%
%
%
%
%
%
%
%
%
%
%
%
%
%
%
%
%
%



\end{thebibliography}
\end{document}